| | |
|---|---|
| Title | Understanding polyethylene surface functionalization by an atmospheric He/O$_2$ plasma through combined experiments and simulations |
| Authors | T Dufour[1], J Minnebo[2], S Abou Rich[1], E C Neyts[2], A Bogaerts[2] and F Reniers[1] |
| Affiliations | [1] Service de Chimie Analytique et de chimie des Interfaces, Faculté des Sciences, Université Libre de Bruxelles, CP-255, Bld du Triomphe, B-1050 Bruxelles, Belgium<br>[2] Research Group PLASMANT, Department of Chemistry, University of Antwerp, Universiteitsplein 1, B-2610 Wilrijk-Antwerp, Belgium |
| Ref. | J. Phys. D: Appl. Phys., 2014, Vol. 47, Issue 22, 224007 (12 pp) |
| DOI | http://dx.doi.org/10.1088/0022-3727/47/22/224007 |
| Abstract | High density polyethylene surfaces were exposed to the atmospheric post-discharge of a radiofrequency plasma torch supplied in helium and oxygen. Dynamic water contact angle measurements were performed to evaluate changes in surface hydrophilicity and angle resolved x-ray photoelectron spectroscopy was carried out to identify the functional groups responsible for wettability changes and to study their subsurface depth profiles, up to 9 nm in depth. The reactions leading to the formation of C–O, C=O and O–C=O groups were simulated by molecular dynamics. These simulations demonstrate that impinging oxygen atoms do not react immediately upon impact but rather remain at or close to the surface before eventually reacting. The simulations also explain the release of gaseous species in the ambient environment as well as the ejection of low molecular weight oxidized materials from the surface. |

# 1. Introduction

The wettability of polymer surfaces is of great importance for various industrial applications: deposition of stable coatings, pattern formation in soft lithography, biocompatibility of coated implants and smart surfaces that are capable of responding to physico-chemical changes of the surrounding medium, such as temperature, pH, [1–3].

It is well established that the wetting properties of polymer surfaces result mostly from chemical interactions and surface roughness, the latter being also considered as mechanical interlocking, in the case of adhesion phenomena [4–8]. In the case of polyethylene, the wettability surface properties are poor due to the low PE surface energy, its chemical inertness and its smooth surface morphology. They can, however, be enhanced by exposing the polymer to a treatment that induces morphological changes by etching and roughening processes, as well as chemical bond cleavage. This last mechanism can lead to: (i) the production of free radicals, which can interact with the environment to produce surface active species, such as polar groups [9]; (ii) the release of low mass stable degradation products, such as the low molecular weight oxidized materials (LMWOM) [10]; (iii) the formation of a large number of double bonds and extensively cross-linked structures.

Among the treatments already used, irradiation treatments (laser, ion beam, electron beam, UV, x-rays and γ-rays) lead to dramatic modification of polymer surfaces in terms of thickness and structure [11]. Svorcik et al [12] induced structural changes of polymer surfaces exposed to an energetic ion beam of 40 keV Ar$^+$ ions. They proved the production of oxidized structures and conjugated double bonds in the surface, determining a significant oxidation depth as high as 120 nm, using infrared, ultraviolet-visible spectroscopies and Rutherford backscattering. However, the fluence of the Ar$^+$ ions had to be limited to prevent the prevalence of deoxygenation processes.

As an alternative, Truica-Larasescu et al [10] managed to incorporate high concentrations of bonded nitrogen and oxygen (almost 10%) on low density polyethylene (LDPE) surfaces by VUV irradiation in ammonia. The surface energy of LDPE was strongly increased, and was even higher than the value obtained in the case of a $N_2$ plasma treatment, although the surface became partially damaged due to significant changes in the surface energy.

In contrast, the treatment of polymer by cold plasmas encountered a great success as the modifications could be applied to a lesser depth and under milder conditions [13, 14]. As a result of a low pressure Ar plasma treatment, Svorcik et al [9] obtained texturized high density polyethylene (HDPE) films with a subsurface oxidation depth of about 25 nm. Carbonyl, carboxyl and amide groups were detected with C=C bonds either in aromatic or in aliphatic groups. Sanchis et al [13] exposed LDPE surfaces to a low pressure $O_2$ plasma and showed an improvement in the surface wettability for short exposure times, by promoting the formation of polar groups (mainly carbonyl, carboxyl and hydroxyl groups). Longer exposure times caused slight abrasion on LDPE films evidenced by a small increase in the surface roughness. By means of attenuated total reflectance Fourier transform infrared spectroscopy (ATR-FTIR) and x-ray photoelectron spectroscopy (XPS), Lehocky et al [14] evidenced the formation of carbonyl, carboxyl, ether and peroxide functionalities resulting from an oxidative (oxygen and air) atmospheric pressure RF-plasma treatment of HDPE.

Using an atmospheric He-$O_2$ post-discharge generated by a radiofrequency (RF) plasma torch can be regarded as a promising approach offering several advantages: (i) the surface properties of the material can be modified without altering its bulk properties (Young modulus, yield strength, etc.), (ii) plasma treatments are environmentally friendly (no need of chemical reactants commonly used in multiple step processes), (iii) they are dry processes, (iv) working at atmospheric pressure prevents the use of expensive pumping systems and (v) working in the post-discharge regime instead of the discharge ensures a milder treatment than usually attained in low pressure discharges. As observed by Abou Rich et al [6] from micrographs obtained by scanning electron microscopy (SEM), LDPE surfaces exposed to a dielectric barrier discharge (DBD) show micro-scale protrusions while the latter appear negligible in the case of an atmospheric post-discharge treatment, for similar plasma conditions.

In the present paper, we study the compositional modifications of the (sub)surface of HDPE exposed to a He or a He-$O_2$ post-discharge by investigating the effects of the treatment time and of the gap (i.e. the distance separating the plasma torch from the polyethylene surface). X-ray photoelectron spectroscopy (XPS—chemical surface composition and chemical bonds) and dynamic water contact angle (WCA—hydrophilicity) measurements are performed. The increase in wetting properties achieved by plasma treatment is typically lost over time due to a slow loss of the surface functionalization until a partial recovery of the native wettability. This so-called ageing effect is also studied here. However, the fundamental interpretation of the reaction mechanisms occurring at the post-discharge/polyethylene interface remains difficult in our case, essentially because of the atmospheric pressure and the interaction between the post-discharge and the environment. The experimental work is therefore complemented by simulations in which we only focus on the main reactive species experimentally evidenced in He–$O_2$ atmospheric flowing post-discharges, namely the oxygen radicals [8, 15]. Atomistic simulations of oxygen atoms impinging on and reacting with a polyethylene surface are carried out, as well as simulations of oxygen atoms already diffused into the (sub)surface layer, in order to highlight the mechanisms occurring at the interface. Furthermore, the plausible reaction pathways between oxygen atoms and polyethylene chains are proposed by calculating their activation barriers with nudged elastic band (NEB) calculations.

## 2. Experimental setup

### 2.1. Material

In this work, large 1mm thick foils of HDPE were purchased from Goodfellow. The foils were cut into 80 × 15mm² sized samples and subsequently cleaned in pure iso-octane and then in pure methanol. After this cleaning procedure, the samples were exposed to the plasma post-discharge.

### 2.2. Plasma source features

The sample surfaces were exposed to the post-discharge of an RF atmospheric plasma torch (AtomfloTM 400 LSeries, SurfX Technologies) [16]. The controller of this plasma source included an RF generator (27.12 MHz), an auto-tuning matching network and a gas delivery system with two mass-flow controllers to regulate the helium and oxygen gases supplying the plasma source. Helium (carrier gas) was used with the flow rate fixed at 15 L.min$^{-1}$, in combination with oxygen (reactive gas) with flow rates ranging from 0 to 200 mL.min$^{-1}$. The RF power coupled into the plasma source was fixed at 120W. As shown in figure 1, the resulting gas mixture enters through a tube connected to a rectangular housing in which the gas was homogenized by crossing two perforated sheets. The gas then flows around the upper electrode and passes through a slit in the centre of the lower electrode. A plasma is struck and maintained between these electrodes by applying an RF power to the upper electrode while the lower electrode is grounded. The slit has an aperture length of 20mm and a width of 0.8 mm. A thorough background characterization of the mechanisms governing this flowing post-discharge has previously been published [15].

A robotic arm was integrated to the plasma torch, enabling the treatment of large samples located downstream. In all our experiments, the plasma source was only moved in one direction along a scan length of 100 mm at a velocity of 25mm.s$^{-1}$. The gap was varied between 1mm and 25 mm.

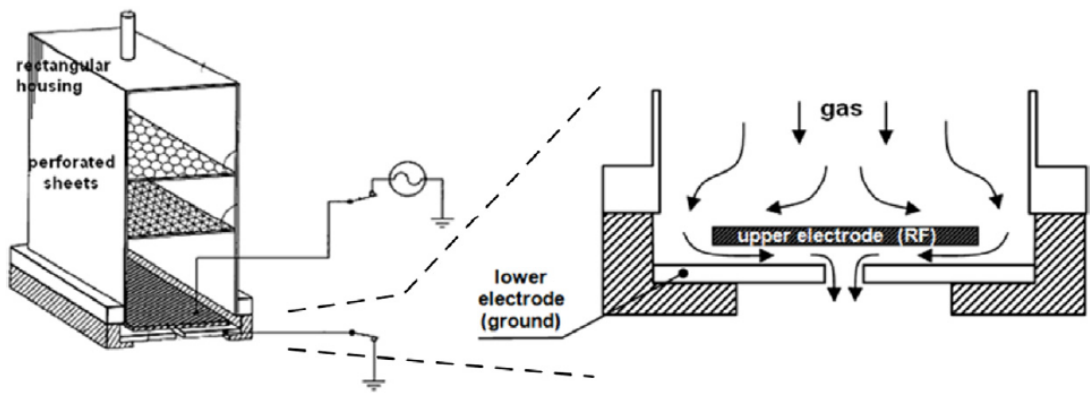

Figure 1. Cross sectional diagram of the RF-plasma source (Patent US7329608 SurfX Technologies [17]), with the entire setup at the left and the detail of the plasma source at the right.

## 2.3. Diagnostics

A drop shape analyser (Krüss DSA 100) was employed to measure dynamic contact angles of water drops deposited onto HDPE samples. Advancing and receding WCAs were measured by growing and shrinking the size of a single drop on the HDPE surface, from 0 to 15µl at a rate of 30µl min−1 [18]. The WCA plotted in this article correspond to the advancing water contact angles (aWCA); rWCA are not plotted for the sake of clarity but also because they are known to be insensitive to the low-energy surface components. Moreover, they are not easily measurable in the case of our plasma-treated surfaces due to the already low values of the advancing WCA [19].

To evaluate the chemical composition at the surface of the samples, XPS analyses were performed (Physical Electronics PHI-5600). The base pressure in the analytical chamber was ≈10⁻⁹ mbar. Spectra were acquired using the Mg anode (Mg Kα line at 1253.6 eV) operating at 300W. Wide spectra were used to determine the chemical elements present at the HDPE surface [20]. Narrow-region spectra were used for the chemical study of the C 1s and O 1s peaks and high resolution spectra were measured to evidence the nature of the bonds between C and O. The values of the pass-energies, number of accumulations, time/step and energy/step parameters are reported in table 1.

The elemental composition was calculated after removal of a Shirley background line and using the sensitivity coefficients: $S_C$ = 0.205 and $S_O$ = 0.63 [21]. The data were acquired with the AugerScan software and the peak fittings were achieved with the Casa XPS software, with FWHM set to 1.5 eV for all the spectral components. The resulting compositions must be taken as indicative and are used only for comparison between the different plasma treatments (with/without $O_2$). They do not reflect the absolute surface composition [22]. In our case, several analysed regions on the same sample always led to the same relative compositions.

| | Wide spectra | Narrow region spectra | High resolution spectra |
|---|---|---|---|
| Pass-energy | 187.85 eV | 187.85 eV | 23.5 eV |
| Accumulations | 2 | 3 | 5 |
| Time/step | 100 ms | 100 ms | 50 ms |
| Energy/step | 0.8 eV | 0.4 eV | 0.05 eV |

*Table 1. Experimental conditions used for the ARXPS measurements*

| | | TOA | | | | | | |
|---|---|---|---|---|---|---|---|---|
| | | 15° | 25° | 35° | 45° | 55° | 65° | 75° |
| z (nm) | O 1s | 2.0 | 3.3 | 4.4 | 5.5 | 6.4 | 7.1 | 7.6 |
| | C 1s | 2.4 | 3.9 | 5.3 | 6.5 | 7.5 | 8.3 | 8.9 |

*Table 2. Sampling depth analysis (nm)*

To carry out the angle resolved x-ray photoelectron spectroscopy (ARXPS) measurements, the take-off angle (TOA), defined between the analyser axis and the surface, was varied by steps: 15°, 25°, 35°, 45°, 55°, 65°, 75°. The angular acceptance and the number of angular channels were 60° and 120°, respectively. In a film of slightly oxidized polyethylene, the photoelectron emitted by an O 1s atom presents a kinetic energy of ≈721 eV corresponding to an attenuation length (λ) of 2.5 nm. In the case of a C 1s atom, this electron kinetic energy is estimated to ≈969 eV, which corresponds to λ = 3.1 nm (23). For TOA = 90°, the average XPS analysis depth from which ≈63% of the signal originates could be calculated by considering a single attenuation length. But it is also well established that 95% of the information obtained by XPS comes from within three attenuation lengths of the surface (3λ). For this reason, the analysis depth (z) is calculated from equation (1), according to which z = 2.8λ for TOA = 75°, while z = 0.8λ for TOA = 15° [24, 25]. Table 2 represents the analysis depths (nm) calculated from equation (1) by considering that the intensity on the top surface (with TOA equal to 15°) is the reference intensity, since the XPS analyser limits are 15° and 75° [26, 27]: $z(nm) = 3 \cdot \lambda \cdot \sin(TOA)$. (1)

## 3. Molecular dynamics and nudged elastic band simulations

Molecular dynamics (MD) is a simulation technique allowing one to follow the time evolution of a set of atoms through space. Interatomic interactions are governed by the interatomic potential used. The negative gradient of this potential yields the forces on the atoms, while their motion is calculated by solving the set of coupled equations of motion using the velocity Verlet algorithm. The simulations in this work utilize the reactive force field (ReaxFF), using parameters for hydrocarbon oxidation, developed by Chenoweth et al [28]. In ReaxFF, several partial energy functions contribute to the total system energy, as given by [29]

$$E = E_{bond} + E_{over} + E_{under} + E_{val} + E_{pen} + E_{tors} + E_{conj} + E_{vdWaals} + E_{Coulomb} \quad (2)$$

These terms account for the bond energy between two connected atoms ($E_{bond}$), the additional energy caused by the over-coordination ($E_{over}$) or under-coordination ($E_{under}$) of an atom, the valence angle energy between triplets of atoms ($E_{val}$), the penalty energy associated with two double bonds sharing an atom in a valency angle ($E_{pen}$), the torsional angle energy between quadruplets of atoms ($E_{tors}$), the energy of conjugation effects in unsaturated systems ($E_{conj}$), the van der Waals interaction energy between all pairs of atoms ($E_{vdWaals}$) and the Coulomb interaction energy between all pairs of atoms ($E_{Coulomb}$), respectively.

The simulations of a polyethylene system were carried out using the LAMMPS simulation package [30]. Periodic boundary conditions were imposed in all directions. The time step was set to 0.25 fs. The polyethylene structures used in the simulations are built from branched chains that extend through the periodic boundaries in the x and y directions. The structure size was set to 30Å × 30Å × 20Å in the x, y and z directions, respectively. The minimum chain length was set to ten carbon atoms, while the branching parameter was chosen such that a side chain of at least one ethylene unit was placed on approximately one out of every ten carbon atoms of the primary chain. Furthermore, a few ether and alcohol functional groups were added to represent a native oxygen content of 1.2 wt%. The final polyethylene structures had a density varying from 0.91 to 0.93 g.cm$^{-3}$ and contained around 2150 atoms. A representative example of a simulated PE structure is shown in figure 2. In the case of subsurface oxidation, the PE structure fills the whole simulation box to represent a portion of the subsurface, whereas in the case of oxygen atom impact, the simulation box is extended by 30Å in the z direction, while the positions of the atoms in the bottom 2Å are kept fixed to mimic the rigidity of the deeper layers. As a result, the PE surface is free to expand into the vacuum at the top of the structure and an interface layer of about 5Å is formed.

Additionally, NEB simulations [31] were performed on a system consisting of a decane molecule and a single oxygen atom, to calculate the potential energy barriers for some plausible reaction pathways. Decane was chosen as a model for polyethylene because (i) it is large enough to represent a single polyethylene chain and (ii) it is small enough to allow reasonable simulation times.

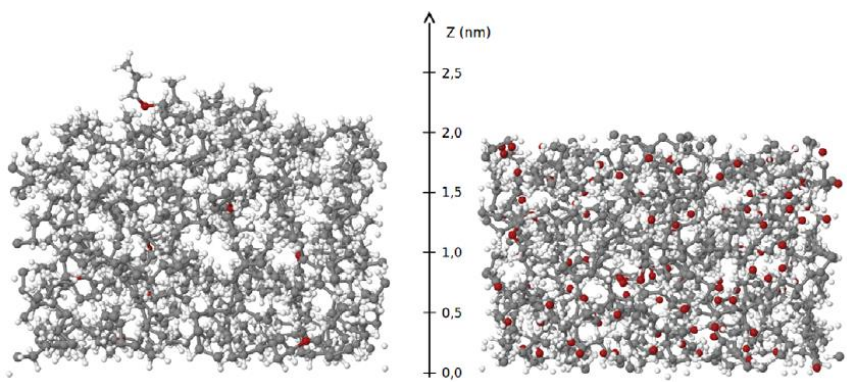

*Figure 2. Side view of the polyethylene structure, as used in the MD simulations for oxygen atom impact (left) and for subsurface layer oxidation (right), with carbon, hydrogen and oxygen atoms represented in grey, white and red, respectively.*

# 4. Experimental results

## 4.1. Influence of the treatment time

The influence of the treatment time on the HDPE samples exposed to the plasma torch can be evaluated in terms of the number of scans. The modifications in surface hydrophilicity as a function of the number of scans are reported in figure 3 for a torch-to-substrate distance (gap) of 1 mm, a scan velocity of 25mm.s$^{-1}$, an RF power of 120W and flow rates of 15 L.min$^{-1}$ and 200 mL.min$^{-1}$ for helium and oxygen, respectively. Whatever the exposure time, the surface hydrophilization seems instantaneous since a single scan enables a decrease of the aWCA from 90° (i.e. native WCA) to 35°. Treatments performed with more scans do not significantly influence the hydrophilicity of the surface since the WCA values remain more or less constant, and always lower than 40°.

This WCA characterization was complemented by XPS measurements. In figure 4, the oxygen surface concentration increases from 20% obtained for a single scan to 32% after five scans, and then remains more or less constant. For comparison, the native HDPE subsurface was slightly oxidized (almost 3%; also indicated in figure 4).

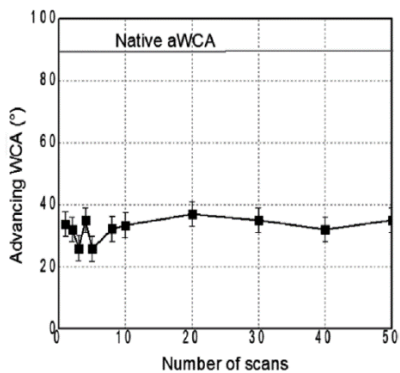
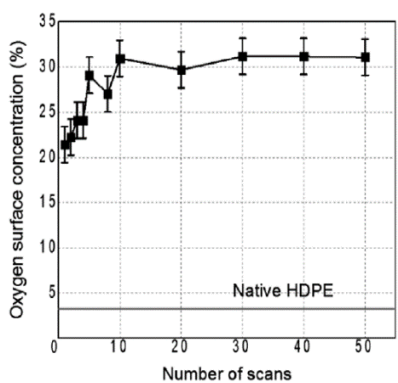
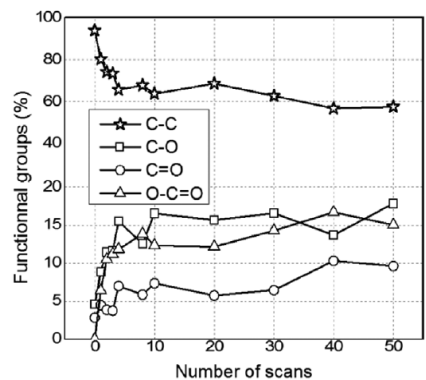

Figure 3. Variation of the aWCA of HDPE surfaces exposed to a He-O$_2$ post-discharge versus the number of scans for $\Phi(He)$ = 15 L.min$^{-1}$, $\Phi(O_2)$ = 200mL.min$^{-1}$, $P_{RF}$ = 120W, gap = 1mm.

Figure 4. Variation of the oxygen surface concentration of HDPE surfaces exposed to a He-O$_2$ post-discharge versus the number of scans for $\Phi(He)$ = 15 L.min$^{-1}$, $\Phi(O_2)$ = 200mL.min$^{-1}$, $P_{RF}$ = 120W, gap = 1mm, TOA = 75°.

Figure 5. Relative content of C–C, C–O, C=O and O–C=O chemical bonds measured by XPS as a function of treatment time expressed as number of scans, with $\Phi(He)$ = 15 L.min$^{-1}$, $\Phi(O_2)$ = 200 mL.min$^{-1}$, $P_{RF}$ = 120W, gap = 1mm, TOA=75°.

To verify whether the oxidation of the HDPE surface corresponds to a functionalization of the surface, the C 1s peak in the XPS measurements was decomposed into several components corresponding to the C–C, C–O (i.e.] ether), C=O (i.e. carbonyl) and O–C=O (i.e. carboxylic) chemical bonds identified at 285.0 eV, 286.5 eV, 288.0 eV and 289.1 eV binding energies [8], respectively. In figure 5, the relative content of these chemical groups is plotted versus the number of scans. The fraction of C–C groups drops from almost 100% to about 65% after ten scans and these bonds are replaced by oxygenated groups which increase from a few percents (<5%) to approximately 10–15% after the same number of scans. Subsequently, the fraction of C–C groups remains more or less constant at a value of 60%, while the fractions of C–O, C=O and C–O=O groups also remain more or less constant with values around 15%, 5–10% and 12–16%, respectively.

## 4.2. Depth profiles of the functional groups

Besides the information about the fractions of these functional groups determined for a TOA fixed at 75°, it is also interesting to investigate their depth profile, in order to evaluate which functional groups are the most anchored onto the HDPE surface. Figure 6 shows the oxygen depth profiles, both for an untreated and a treated HDPE sample (after five scans). In the case of the untreated (native) HDPE surface, the oxygen concentration drops from almost 5% to 3% in the first 8 nm, while for the treated surface, the oxygen concentration decreases from 33% to 29%. The main feature is obviously a significant difference in the O-concentration between the treated and the untreated sample. However, in both cases, the variations in relative O-concentration between 0 and 8 nm remain lower than 5%. These very low variations may be attributed to a too high $O_2$ flow rate (200 mL.min$^{-1}$), which immediately saturates the subsurface of the polymer, although experiments performed with a flow rate of only 20 mL.min$^{-1}$ give the same results. An $O_2$ flow rate comprised between 1 and 5mL.min$^{-1}$ may be required to observe significant variations in O-concentration but they cannot be experimentally obtained since the mass flow meters integrated to the plasma source present a too low accuracy (10 mL. min$^{-1}$ in step).

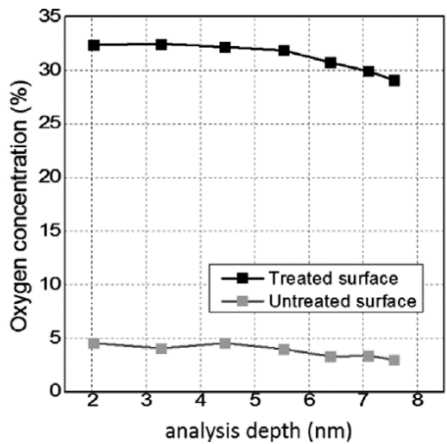
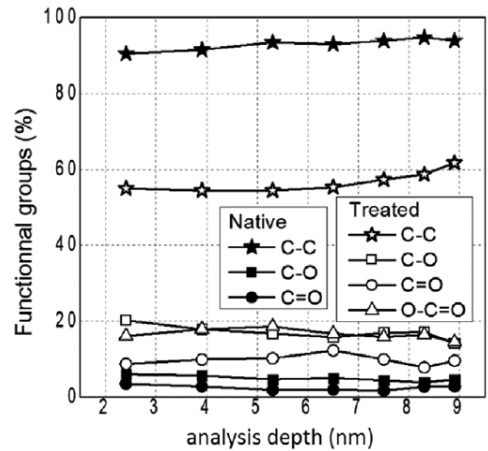

Figure 6. Depth profile of the oxygen concentration in the subsurface of both an untreated HDPE sample, and a treated sample (after five scans), with the same experimental conditions as reported in figure 5.

Figure 7. Depth profiles of the C–C, C–O, C=O and O–C=O chemical bonds for a native and a plasma-treated HDPE surface (after five scans), with the same experimental conditions as reported in figure 5.

In figure 7, the depth profiles of the functional groups are plotted for the same samples as in figure 6. The profiles of the oxygenated functional groups from the topmost surface until a depth of 9 nm remain almost constant while presenting some local variations. The latter ones could result from randomly oriented distributions of the oxygenated functions. In the untreated sample, the dominant fraction is the C–C bonds (90% and more), while the remaining fraction (<10%) is attributed to C–O and C=O bonds. For the treated surface, the fraction of C–C bonds is still around 55–60%, but the C–O, O–C=O and C=O bonds are now also clearly visible. The most important of these functional groups are C–O and O–C=O (slightly below 20%), while the C=O bonds have a lower concentration of around 10%, analogous to the results shown in figure 5 above.

## 4.3. Influence of the gap

The gap (i.e. the distance separating the plasma torch from the polyethylene surface located downstream) is a critical parameter for the treatment of polymer surfaces with a plasma torch. On one hand, if the gap is too small, the polymer surface can reach its melting point in the case of a long treatment time due to the post-discharge heating. On the other hand, if the gap is too large, the interaction of the post-discharge with the environment air is enhanced, thus giving rise to new reactions with new impurity sources, with the risk of limiting the treatment efficiency. Figure 8 indeed shows that the advancing WCA values increase from about 30–40° for a gap of 1 mm, to values around 60°–80° when the gap rises to 25 mm. Hence, they approach the threshold of 90°, corresponding to the advancing WCA of a native HDPE surface. This indicates indeed that a larger gap reduces the treatment efficiency.

Furthermore, it is clear from figure 8 that the aWCA values depend on whether oxygen is mixed with the carrier gas (helium) or not. The addition of 200 mL.min$^{-1}$ of oxygen to the 15 l min$^{-1}$ of helium gas represents a variation of only 1.3% in the global gas flow rate, but the differences in aWCA values measured at a gap of 1mmare about 10° (i.e. 30° versus 40°), and the difference becomes even more significant at a gap of 25 mm, i.e. 82° (without oxygen) compared to 66° (with oxygen). Hence, the treatment with a small fraction of oxygen in the helium gas flow gives rise to clearly lower aWCA values, indicating that more oxygen-containing functional groups are built in, as expected. Plotting the variations in oxygen surface concentration versus the gap is therefore relevant to verify if this surface hydrophilization is correlated with the grafting of oxygenated functional groups.

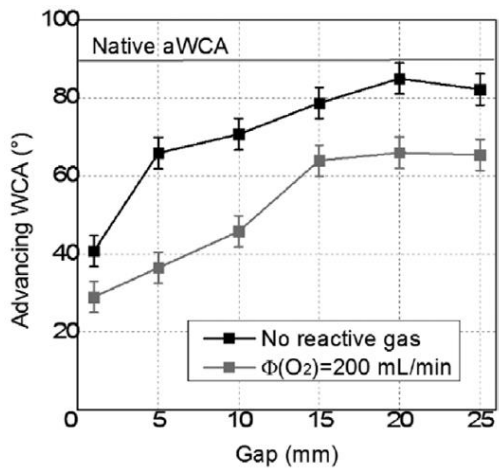
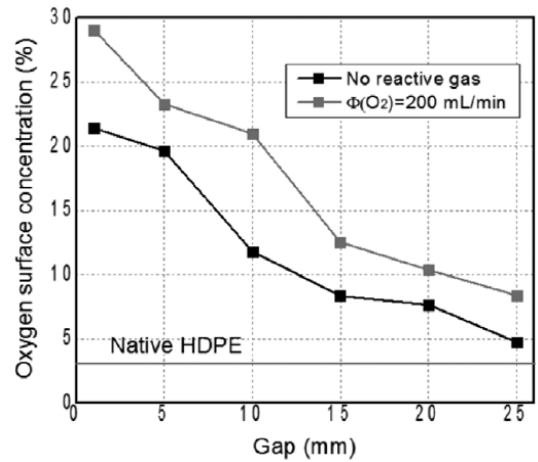

Figure 8. Variation of the aWCA of HDPE surfaces exposed to a He post-discharge, with and without the addition of oxygen, versus the gap between 1mm and 25 mm, for $\Phi$(He) = 15 L.min$^{-1}$, $\Phi$(O$_2$) = 0 or 200 ml min$^{-1}$, $P_{RF}$ = 120W, 5 scans.

Figure 9. Variation of the oxygen surface concentration of HDPE surfaces exposed to a He post-discharge, with and without the addition of oxygen, versus the gap between 1 and 25 mm, for $\Phi$(He) = 15 L.min$^{-1}$, $\Phi$(O$_2$) = 0 or 200 mL.min$^{-1}$, $P_{RF}$ = 120W, 5 scans, TOA = 75°.

Figure 9 indicates a significant drop in the oxygen concentration when the gap is increased from 1 to 25 mm. Indeed, in the case of a He-O$_2$ post-discharge, the oxygen concentration drops from about 30% at a gap of 1 mm, towards 8% at a gap of 25 mm; in the pure He post-discharge, the oxygen concentration drops from 21% to 5%, hence approaching the value of 3%, characteristic for the native HDPE surface. This indicates again that the treatment becomes less efficient for larger gaps. The fact that the He–O$_2$ treated surface shows higher oxygen concentrations is logical, and consistent with the results from figure 8.

The depth profiles of the C–C, C–O, C=O and O–C=O functional groups are plotted in figure 10 for a gap of 1mm and 10 mm. The fractions of these functional groups remain again more or less constant as a function of depth, as was also illustrated in figure 7. Furthermore, when comparing the results obtained for the two different gaps, it is clear that the fraction of the C–O and C=O groups are fairly independent of the gap (i.e. around 15–20% and 7–10%, respectively), while the formation of O–C=O groups seems to be enhanced for the smallest gap (1 mm); as a consequence, the fraction of C–C bonds is somewhat lower in the case of a 1mm gap.

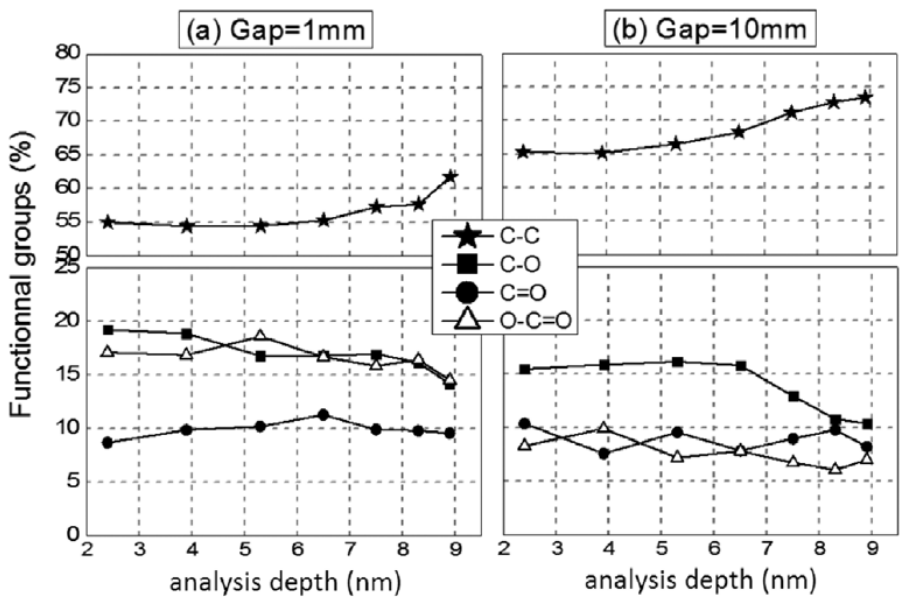

*Figure 10. Depth profile of the C–C, C–O, C=O and O–C=O chemical bonds for plasma-treated HDPE surfaces with a gap of (a) 1mm and (b) 10 mm. In both cases, $\Phi(He) = 15$ L.min$^{-1}$, $\Phi(O_2) = 200$ mL.min$^{-1}$, $P_{RF} = 120W$, 5 scans.*

## 4.4. Effect of ageing time

It is clear that the atmospheric plasma treatment increases the amount of polar groups on the polyethylene surface, but in order to determine whether this increase is irreversible or not, a dedicated study was carried out to investigate the evolution of the surface wettability over time. The plasma-treated films were subjected to ambient air ageing (70% relative humidity (RH) and 293 K) to evaluate their hydrophobic recovery time. The samples were not rinsed after plasma treatment and before WCA measurements. Their chemical compositions are plotted as a function of depth in the surface in figure 11 for a freshly treated film and a film treated one month ago. Figure 11 shows that the oxygen concentration is fairly constant as a function of depth, as is also illustrated above in figure 6, but a decay in the oxygen concentration from 32% to 24% is observed after one month, indicating that the surface hydrophilicity slightly drops again after the plasma treatment.

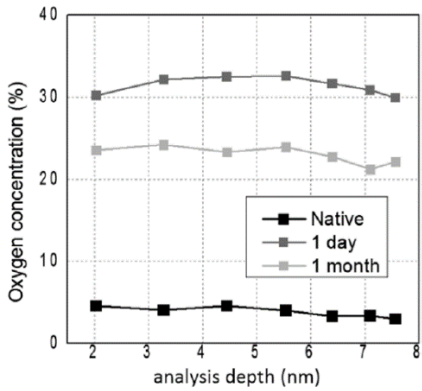

Figure 11. Chemical surface composition of a HDPE sample obtained by XPS the same day as its plasma treatment ($\Phi(He)$ = 15 L.min$^{-1}$, $\Phi(O_2)$ = 200mL.min$^{-1}$, $P_{RF}$ = 120W, 5 scans, gap = 1 mm) and one month later.

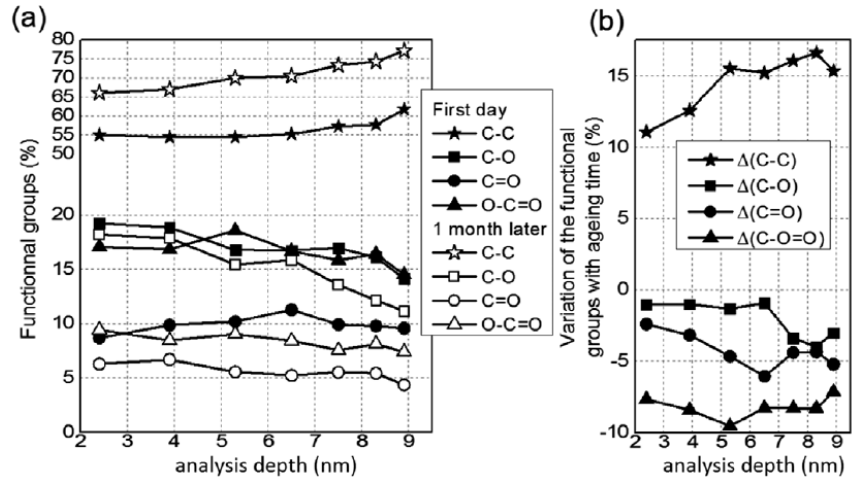

Figure 12. (a) Depth profiles of the functional groups measured the same day as the plasma treatment and one month later. (b) Variations in the depth profiles over one month. The experimental conditions are the same as in figure 11.

In figure 12(a), the fractions of the functional groups are plotted as a function of depth, for both the freshly treated film and the film treated one month earlier. It is clear that the fraction of C–C bonds increases, whereas the fractions of O containing bonds decrease with ageing time, but to a different extent. For example, at a depth of 2.4 nm, the fraction of C–C groups increases from 55% to 66%, the fractions of C–O and C=O groups drops only slightly from 19% and 10% to about 18% and 6% respectively, while the fraction of O–C=O groups decreases more significantly from 17% to 9%. For the sake of clarity, the variations in the fraction of functional groups are also plotted in figure 12(b) versus the depth as measured by XPS. Hence, it is clear that the fractions of C–O and C=O groups drop only by a few per cent (to maximum 5%) over one month, irrespective of the depth, while the fraction of O–C=O groups shows a decrease between 7% and 10% on the whole subsurface, and the fraction of C–C groups rises considerably, especially at larger depths (i.e. around 10% at the 2mm depth, but up to 17% at the 8mm depth).

## 5. Calculation results

According to the experiments, the He-O$_2$ plasma treatment introduces a significant amount of oxygen functional groups on the HDPE surfaces. However, the processes contributing to this result are less evident. Therefore, MD simulations and NEB calculations were performed to obtain a better understanding of the interactions between polyethylene and oxygen atoms, which are assumed to be the main plasma species responsible for surface oxidation.

### 5.1. MD simulations

First, the reactivity of oxygen atoms upon impact with the polyethylene surface was investigated. The oxygen atoms impinge on the surface with a thermal energy of 0.025 eV, which is a realistic assumption for a post-discharge setup as used in the experiments. In these simulations, the oxygen atoms do not react immediately upon impact with a polyethylene surface (i.e. within the simulation time of 2 ps). Almost 90% of the impinging oxygen atoms, however, are not reflected from the surface either, but remain at the surface or diffuse slightly into it. These results indicate that the oxygen atoms react only some time after they reach the surface and that they can readily diffuse into the surface layer before reacting.

Second, we also studied the reactivity of oxygen atoms in the subsurface region. Two different polyethylene structures, representing the subsurface, were modified by placing oxygen atoms into them, representing 10 and 15 wt% oxygen. Within a simulation time of 2 ns at 300 K, 7% and 17% of the oxygen atoms had reacted with the polyethylene for the two different polyethylene structures, respectively, while 47 and 36% had recombined to form oxygen molecules. However, the formation of new functional groups, as evidenced by the XPS measurements, was not observed within this short timescale. It is clear from the simulations that the reactivity of the polyethylene increases significantly at the site of an initial bond breaking reaction. This leads to the hypothesis that the activation energy for the first bond breaking reaction is higher than for subsequent reactions.

## 5.2. NEB calculations

In order to verify this hypothesis and to predict which reaction pathways are important for the introduction of functional groups to the polyethylene surface, NEB calculations were performed to calculate the activation energy of the most plausible reactions between an oxygen atom and a decane molecule (see figure 13).

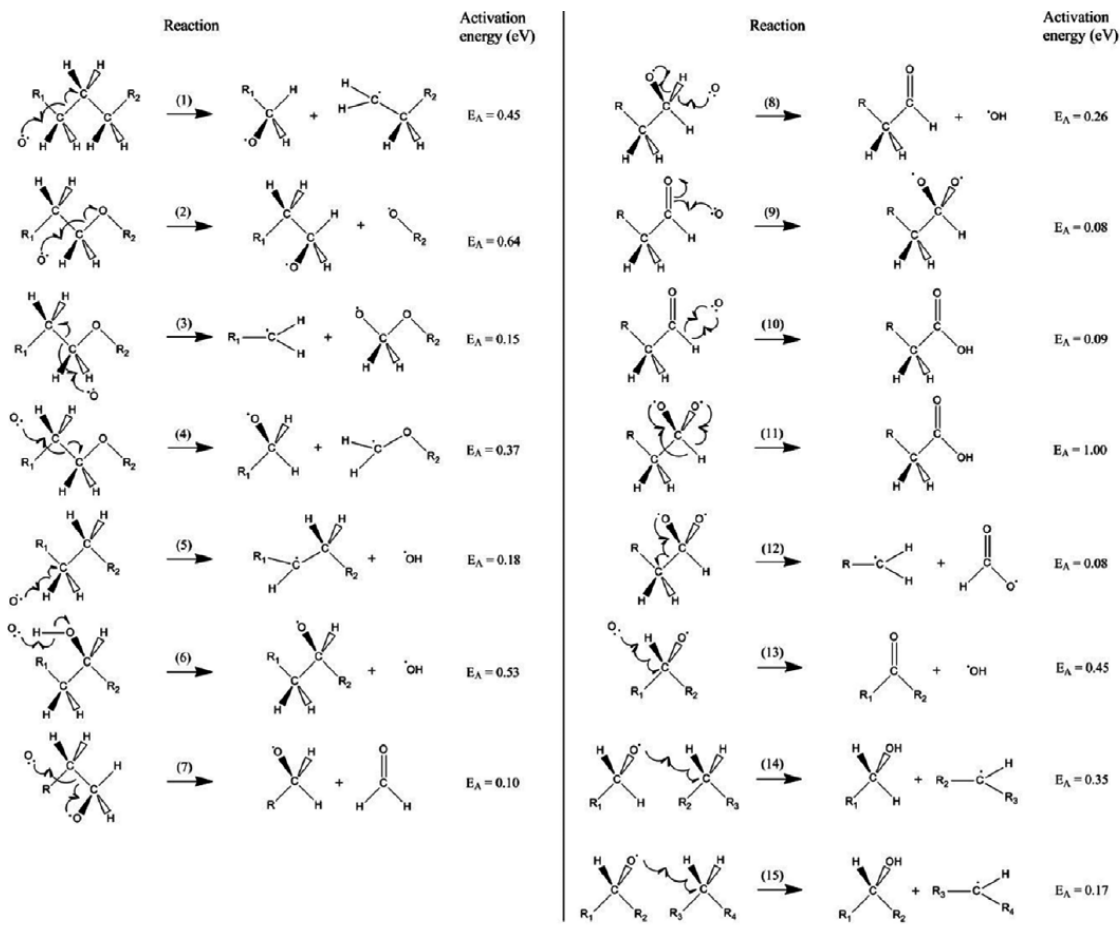

*Figure 13. Activation energies in eV for plausible reaction pathways between an oxygen atom and a decane molecule or a derivated compound.*

Reaction (1) depicts the oxidative cleavage of a carbon-carbon bond in a pristine polyethylene chain, reaction (2) illustrates the carbon–oxygen bond breaking of an ether group, and reactions (3) and (4) show the carbon–carbon bond breaking near an ether group. These reactions, in combination with the hydrogen abstraction reactions (5) and (6) at the polyethylene chain or at an alcohol group, respectively, are considered as the initial reactions in this work. Subsequently, reactions (7)–(15) can take place. In reaction (7), a formaldehyde molecule is split from the chain by a carbon–carbon bond breaking event, while in reactions (8) and (13), an aldehyde and a ketone, respectively, are formed by a hydrogen abstraction. Further reactions of ketones are not shown in figure 13, as the activation energies are similar to those of aldehydes and moreover the formation of a ketone was not observed during the simulations. When an aldehyde is produced, it can react with another oxygen atom to form a carboxylic acid, as shown in a single step in reaction (10), or in two steps in reactions (9) and (11). However, the intermediate product formed in reaction (9) is not very stable and can easily dissociate, as depicted in reaction (12), to split the oxidized carbon atom from the chain. Finally, reactions (14) and (15) depict the hydrogen abstraction from a second decane molecule, by an oxygen radical attached at the end or in the middle of the chain, respectively, to form an alcohol.

The values obtained from the NEB calculations confirm that an initial carbon–carbon bond breaking reaction increases the reactivity of the polyethylene chain at that site. For example, the potential energy barriers for the first and second carbon–carbon bond breaking (reactions (1) and (7)) are 0.45 eV and 0.10 eV, respectively. When comparing the potential energy barrier for reaction (1) with the barrier obtained from discrete Fourier transform (DFT) calculations [32], our calculated energy barrier is quite low (0.45 versus 1.95 eV), but the relative order of the energy barriers, at least for reactions (1) and (5), was found to be consistent with DFT simulations. For the other reactions, no DFT calculations could be found in the literature. From the calculated activation energy and the chemical structure of polyethylene, it is expected that the most important initiation reaction is the hydrogen abstraction (reaction (5)). However, this reaction was only observed a single time during the MD simulations. Instead, the carbon–carbon bond breaking (reaction (1)) was observed as the prevalent initiation reaction. Since NEB calculations are static simulations and MD simulations are dynamic, this discrepancy in reactivity can be attributed to the dynamic steric hindrance for the hydrogen abstraction, due to the rotation of the methylene group, away from the approaching oxygen atom.

For subsequent reactions, the activation energy is lower for a second carbon–carbon bond breaking (reaction (7)) than for a hydrogen abstraction at the activated site (reaction (8)). In the MD simulations, reaction (7) was indeed found to occur more often, with the result that a formaldehyde molecule was split from the chain, which was then further oxidized to carbon dioxide. When reaction (8) did occur, the aldehyde was quickly further oxidized through the addition of another oxygen atom (reaction (9)) and subsequently split from the chain via reaction (12), resulting in the loss of the functional group. According to the calculated activation energies, the aldehyde could as well be oxidized to a carboxylic acid (reaction (10)), but this reaction mechanism was not observed in the MD simulations, which may have been due to the limited timescale.

Finally, the formation of alcohol groups was not observed through the proposed reaction mechanisms (15) and (16) involving a hydrogen abstraction, despite their relatively low activation energy. This can (in part) be explained by the fact that these reactions involve two decane molecules, which represent two different polyethylene chains. As such, the probability for these reactions to occur is low due to geometrical constraints on the reactive oxygen radical, which thus imposes an entropic barrier on this reaction.

## 6. Combining the results from experiments and simulations

The results from the MD simulations and NEB calculations correlate very well with the experimental results presented in figure 5. The initial reactions (1)–(6) between oxygen atoms and polyethylene are the main production channels of either hydroxyl radicals (which can either recombine with other radicals or be released to the ambient atmosphere (reactions (5) and (6))), or oxygen radicals, which are connected with a single bond to the polyethylene chain (reactions (1), (2), (3), (4)).

Such a C–O bond is indeed the most abundant functional group according to the XPS measurements (see figure 5). Since the MD simulations have shown that the oxygen radicals are almost exclusively located at the end of a chain, reactions (7) and (8) seem the most important for further oxidation, both leading to the formation of a carbonyl bond. One could therefore expect the increase of the carbonyl bonds to be correlated with a decrease of the ether groups. This is however not observed in figure 5, since the plasma torch continuously provides O radicals to the polyethylene chains, thus simultaneously producing ether groups according to reactions (1)–(4). In the case of reaction (7), the formaldehyde molecule that is produced can be further oxidized to a volatile carbon dioxide molecule in multiple steps, as observed in the MD simulations (as explained above), and disappear from the surface. In the case of reaction (8), the formed aldehyde is not very stable in the oxidative environment during plasma treatment and can react with another oxygen atom, leading to the formation of a carboxylic acid as suggested in reaction (10), or again to the loss of the functional group (reactions (9) and (12)). Therefore, the very low activation energy of reaction (10) of 0.09 eV can explain why the XPS measurements show more O–C=O (i.e. carboxyl) than C=O (carbonyl) groups in figure 5. Indeed, most of the carbonyl groups are expected to be aldehydes, which are likely to be further oxidized to carboxyl groups. Reaction (11) could also explain the formation of carboxyl groups, which however depends on reactions (8) and (9), but the associated activation energy is quite high (1.00 eV).

This explanation also accounts for the large difference in the fraction of carboxyl groups when varying the gap, as observed in figure 10. When the gap is reduced from 10mm to 1 mm, the amount of oxygen atoms that reach the surface increases, meaning that the surface oxidation also increases. This in turn gives rise to a higher reaction rate of all the reactions that require a new oxygen atom, of which the most important for the surface composition (i.e. for the production of a new functional group) are reactions (1), (8) and (10). The formation of ketones by reaction (13) is less important because most of the oxygen radicals are located at the end of a chain, and the formation of carboxylic groups by reaction (11) is not likely because of its high activation energy. If we assume that these three reactions are all dependent in first order on the concentration of oxygen atoms, the fraction of C–O and C=O bonds will remain constant, while the fraction of O–C=O bonds will increase when the gap is reduced, which is exactly what was measured experimentally (see figure 10).

Another correlation between experiment and simulation can be drawn from the ageing study presented in figure 12, albeit less straightforward. A mechanism commonly invoked to explain the hydrophobic recovery over time is a reorientation of the polar groups into the bulk of the polymer. According to Gerenser [33], LMWOMs coming from chain scissions during the plasma treatment could slowly diffuse into the bulk material upon the ageing time, thus contributing to a decrease of the concentration of oxygen on the topmost layers of the film. However, figure 12 shows that the loss of functional groups is almost uniform in depth. This suggests that the ageing does not take place by a rearrangement of functional groups from the surface to the bulk, because in that case, the concentration of functional groups should decrease in the first few nm, in favour of the deeper layers. Instead, the concentration of functional groups decreases more in the deeper layers than in the top layers of the surface. This measurement is (strong) evidence for an upward movement of the functional groups, which can be explained by the creation of LMWOMs during the plasma treatment that do not diffuse in depth. These small fragments are likely to be volatile, and thus will migrate to the surface and subsequently be lost to the environment. This result can be correlated with the works of Payner [34] on polystyrene surfaces exposed to an Ar-$O_2$ plasma, where added oxygen to the surface during the plasma treatment appears not to diffuse into the bulk over time (92 h). Also, the work of Weon [35] evidences the depletion of antioxidant originally added to LDPE after a thermal ageing of 6000 h.

Our explanation for the ageing process is supported by the simulations. As mentioned above, reactions (7) and (12) were frequently observed during the MD simulations, leading directly to the formation of LMWOMs, containing a carbonyl and a carboxyl group, respectively. These fragments can remain trapped in the polyethylene subsurface and be slowly released over time. Indeed, figure 12 shows a greater loss for carbonyl and carboxyl groups, as opposed to alcohol or ether bonds. This line of thought also applies to fragments consisting of several carbon atoms, which are the result of multiple bond breaking reactions on the same polyethylene chain. Such fragments are likely to contain oxygen functional groups at the ends of the fragment, which will most likely be carbonyl or carboxyl groups as a result of reactions (8) and (10), respectively.

# 7. Conclusions

The treatment of HPDE samples by a helium-oxygen post-discharge leads to a higher oxygen concentration in the (sub)surface of the sample and therefore also in higher concentrations of oxygenated functional groups, namely C–O, C=O and O–C=O. This functionalization was studied by ARXPS to determine their depth profiles. In the case of an $O_2$ flow rate of 200 mL.min$^{-1}$, the treatment time only slightly influences the depth profiles while they are more affected by a gap variation from 1 to 25 mm. The ageing study showed that while the concentration of ether groups remains constant, while the concentration of carbonyl and carboxylic groups drops. The formation of low molecular weight oxidized materials such as formaldehyde could explain the instability of the surface over time, while the rearrangement of chains at the surface was found to be unlikely, based on the ARXPS measurements.

Surface analysis (XPS and WCA) experiments and simulations showed a good correlation, and allow one to further understand the mechanisms occurring at the interface between the post-discharge and the polymer. The reactions that were observed in the MD simulations, together with their activation energies as calculated by NEB simulations, can be used to explain the differences in concentration between the different functional groups as measured by XPS. The reactions which are of particular importance for the formation of functional groups at the polyethylene surface are the C–C bond breaking (reaction (1)), the hydrogen abstraction to form an aldehyde (reaction (8)) and the oxidation of an aldehyde to a carboxylic acid (reaction (10)).

# 8. Acknowledgments

This work was part of the IAP/7 (Inter-university Attraction Pole) program 'PSI-Physical Chemistry of Plasma Surface Interactions', financially supported by the Belgian Federal Office for Science Policy (BELSPO). This work was carried out in part using the Turing HPC infrastructure at the CalcUA core facility of the Universiteit Antwerpen (UA), a division of the Flemish Supercomputer Center VSC, funded by the Hercules Foundation, the Flemish Government (department EWI) and the UA.